\documentclass[11pt,a4paper]{article} 
\usepackage{amsmath,amsfonts,amssymb}  
\usepackage{graphicx}
\newcommand{\np}{\nonumber\\}

\title{
Spinons in Magnetic Chains of Arbitrary Spins at Finite Temperatures
}
\author{ J. Suzuki\thanks{e-mail: suz@hep1.c.u-tokyo.ac.jp}\\
        \parbox{0.9\textwidth}{
        {\em
        \begin{center}
       Institute of Physics,\\
        University of Tokyo at Komaba,\\
       Komaba 3-8-1, Meguro-ku, Tokyo,\\
       Japan
        \end{center}
        }}
       }
\begin{document}
\maketitle
\begin{abstract}

  The thermodynamics of solvable isotropic chains with arbitrary spins 
  is addressed by the recently developed quantum 
  transfer matrix (QTM) approach.
  The set of nonlinear equations which exactly characterize
  the free energy is derived by respecting the physical excitations
  at $T=0$, spinons and RSOS kinks.
  We argue the implication of the present formulation to spinon
  character formula of level $k=2S$ SU(2) WZWN model .
  \end{abstract}
\clearpage
%
\section{Introduction}
The 1D spin systems have been providing 
problems of both physical and mathematical 
interests.
Among them, there exists a family of solvable 
models of Heisenberg type with spin-$S$ \cite{Tak,Bab}.
This includes, for instance,
\begin{eqnarray}
{\cal H} &=& J
 \sum_{i=1}^L \{\vec{S}_i \vec{S}_{i+1}+ \frac{1}{4} \}   
  \label{Hamilspinhalf} \\
{\cal H} &=& \frac{J}{4}\sum_{i=1}^L 
   \{\vec{S}_i \vec{S}_{i+1}-(\vec{S}_i \vec{S}_{i+1})^2 +3 \}   
 \label{Hamilspin1}  \np
\end{eqnarray}
as $S=1/2, 1$, respectively.

Ground state properties as well as low lying excitations have
been elucidated by the powerful machinery of solvable models,
the Bethe Ansatz Equation (BAE).
It has been demonstrated in many contexts \cite{Affleck, AGZ, AlcMart}
that the underlying field theory 
is level $k=2S$ SU(2) WZWN model \cite{Witten}.

Although this 1D quantum model is equivalent to a 2D vertex model,
physical excitations carry both natures of
 vertex models and restricted SOS models.
This is firstly demonstrated in \cite{ReshetS} based 
on the $S-$ matrix argument.
The space of states is identified in \cite{ITIJMN}.
By decomposition of crystals, such double feature is
made explicit in terms of "type of domain walls" and "type of domain".
There are independent justifications of this:
The double feature in the spectral decomposition is shown
by path space approach\cite{Nagoya} .
See also \cite{HKKOTY} for 
the decomposition of space picture realized in
Fermionic forms.

Here we are interested in the finite temperature problem.
Standard arguments employ the string hypothesis\cite{Bab, BabTsv}.
The excitation is described 
not in terms of "physical excitations" in the above sense,
but in the "string basis".
There strings  of arbitrary lengths are allowed, which 
results into infinitely many coupled integral equations
among infinitely many unknown functions.
The description successfully reproduces the expected
specific heat anomaly.
It may not be, however,  best suitable for practical
numerics.

We revisit the problem via the recently developed
commuting Quantum Transfer Matrix (QTM) approach
\cite{KluZeit}-\cite{JSE8}.
The formulation does not reply on the string hypothesis.
Rather, it only relies on analyticity structures of the
object called QTM\cite{MSPRB, KluRSOS}.
The problem of the combinatorial summation, i.e., evaluation
of the partition function, then reduces to
investigations on the analyticity of suitably chosen
auxiliary functions.
Up to now, two kinds of choices are adopted independently.

\noindent(A)
The eigenvalue of the QTM as given by the Quantum Inverse Scattering Method
consists of several terms.
The auxiliary functions are chosen from 
combinations of products of these terms
\cite{KluZeit, JKStJ, JKS2p, JKSHub}.
A convenient choice leads to a finite number of 
coupled non-linear integral equations
for a finite number of unknown functions

\noindent (B)
A set of auxiliary functions may be chosen from the
fusion hierarchy among "generalized" QTMs
\cite{KluRSOS, JKSFusion, KSS, JSE8}. 
Generically, one obtains an infinite number of 
coupled non-linear integral equations
for an infinite number of unknown functions.
This can be shown to recover the conventional TBAs based on the string 
hypothesis. Of course, the new approach is entirely free of
any assumption about excitations like string hypothesis.

The spin-1 case is analyzed in the related problem, 
 in the context of finite size corrections\cite{KBP}.
There six functions are introduced in the spirit of (A).
The structure of NLIE among them 
is much more involved in comparison to the spin-1/2 case,
and seems to defy a simple-minded generalization to higher $S$ cases.
In a sense the most subtle point in the QTM approach appears;
one does not know a priori ``better'' set of auxiliary functions.

Here, we adopt other choice of auxiliary functions,
in particular  for the spin 1 case the number of these 
functions is 3 in contrast to 6 as in \cite{KBP}.
A simple idea of combining the two formulations (A) and (B)
works well so that the generalization to
arbitrary $S$ is possible.
The adopted functions agree with the picture 
in \cite{ReshetS}.
Roughly speaking, the fusion part (B) of the auxiliary 
functions is related
to the RSOS piece of the excitation, while  auxiliary functions from
(A) correspond to spinons.

We remark that the 
fusion hierarchy itself is not truncated, by brute force, into finite set.
Instead, the spinon part makes the functional relations 
among them strictly closed.
Thus we obtain $2S+1$ coupled integral equations for $2S+1$ unknown
functions.

Besides the practical advantage, it implies 
the universality in the description of 
thermodynamics of solvable quantum 1D
chains, i.e., 
the description
only in terms of objects which reduce to physical 
excitations in $T \rightarrow 0$.
This has been already demonstrated for several models
in highly correlated 1D electron systems including
the supersymmetric $t-J$ model \cite{JKStJ},
the supersymmetric extended Hubbard model \cite{JKS2p}
and the Hubbard model \cite{JKSHub}.
There the exact thermodynamics are formulated in terms of
"spinons" and "holons", although they lose sense at sufficiently
high temperatures.
The present study adds one successful example 
even in the fusion models 
and gives further supports on the above conjecture.

This paper is organized as follows.
In the next section, we define the main object in this approach, 
the quantum transfer matrix (QTM).
A minimal information of the novel approach is sketched.
Section 3 is devoted to a brief description of the fusion
hierarchy of generalized QTMs.
After these preparations, we introduce auxiliary functions
and examine functional relations among them in Section 4.
The analytic structure studied numerically leads to 
nonlinear integral equations as discussed in Section 5.
Based on these equations
the low temperature asymptotics is studied analytically
in Section 6.
The central charge of the level $k=2S$ SU(2) WZWN model is 
successfully recovered.
We also present the numerical evaluation of specific heat
of $S=1/2, 1, 3/2 $ models for wider ranges of temperatures.
In Section 7, the implication of the present formulation to 
the spinon character formula of the WZWN model \cite{BLS,Nagoya,NakYam}
is discussed.
The summary of the paper is given in Section 8.
%
%
\section{QTM Formulation}

The present QTM formulation originates from
two ingredients:
the equivalence theorem between 1D quantum and 2D classical systems
\cite{MSPRB}
on one side and the integrability structure 
on the other\cite{Baxbook}.
Especially, the latter provides the way to introduce
commuting QTMs 
which reduce the problem of combinatorial counting
to that of analyticity of suitable auxiliary functions\cite{KluRSOS}.
Such a strategy has been successfully applied to several 
interesting models
\cite{KluRSOS}-\cite{JSE8}.
We also mention earlier studies on
thermodynamics \cite{Koma}-\cite{Mizuta}
 which essentially utilize only the former part of ideas. 

A classical analogue to solvable spin-$S$ XXX model
is already found as  a $2S+1$ state vertex model \cite{SAA}.
The Boltzmann weights are identified with the matrix elements
of $\widehat{\mathfrak{sl}_2}$ invariant $R^{\vee}$ matrix:

\begin{eqnarray}
R^{\vee}(u) &=&\sum_{j=0}^k \rho_{2k-2j}(u) P_{2k-2j}  \np
 \rho_{2k-2j} &=& \prod_{\ell=0}^{j-1} \frac{2(k-\ell)-u}{2(k-\ell)}
                  \prod_{\ell=j}^{k-1} \frac{2(k-\ell)+u}{2(k-\ell)}
    \label{Rmat}
\end{eqnarray}
where $k=2S$ and $P_j$ is the projector to 
$V_j$, the $j+1$ dimensional irreducible module of $sl_2$.
We choose $\{-j/2, -j/2+1, \cdots, j/2 \} $ as basis for $V_j$.
The spectral parameter $u$ represents the
anisotropy of the vertex weights.
The  Yang-Baxter equation
implies the  commutation  of row-to row transfer  matrices  for
arbitrary spectral parameters   $u$,   $v$:   
\mbox{${\cal   T}(u){\cal  T}(v)={\cal T}(v){\cal T}(u)$} with

\begin{equation}
  {\cal T}^\beta_\alpha(u)=\sum_\mu\prod_{i=1}^L
  {\cal R}_{\alpha_i\beta_i}^{\mu_i\mu_{i+1}}(u),
\label{transfer}
\end{equation}

where $L$ denotes the real system size, 
$\alpha_i, \beta_i, \mu_i\in V_k$, 
${\cal R}= P R^{\vee}$ and $P(x\otimes y) = y\otimes x$.

\begin{figure}[tbp]
  \begin{center}
    \includegraphics[width=0.20\textwidth]{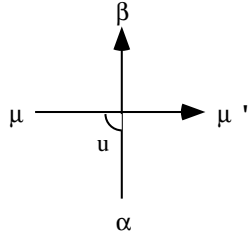}
    \caption{ Graphical representation of the R matrix  ( its element 
     ${\cal R}_{\alpha \beta}^{\mu \mu'})$. 
	  }
	  \label{Rmatfig}
 \end{center}
\end{figure}
The Hamiltonian is 
obtained  as the logarithmic derivative at  
$u=0$,
\begin{equation}
  {\cal H}= J \frac{{\rm d}}{{\rm d}u}
  \ln {\cal T}(u){\Big|_{u=0}}.
\label{logderi}
\end{equation}
It is an easy exercise to verify (\ref{logderi}) gives
(\ref{Hamilspinhalf}) and( \ref{Hamilspin1}) for $S=1/2, 1$ 
respectively.
This may be most easily done by representing $P_j$ in
(\ref{Rmat}) by
$$
P_j = \prod_{p=0, p\ne j}^k \frac{\vec{S} \vec{S} -x_p}
               {x_j-x_p},
$$
and $x_j = 1/2 (\frac{j}{2}(\frac{j}{2}+1) - k(\frac{k}{2}+1))$.
The Hamiltonian for general $S$ can be extracted  similarly.
This is the well-known expression of the equivalence between
1D quantum systems and 2D classical models.
To utilize the equivalence in evaluating finite $T$ quantities,
especially, free energy, 
we need to proceed further.

Let us introduce  Boltzmann weights
$\widetilde{{\cal R}}$ ($\overline{{\cal R}}$   ) of 
 models related  to (\ref{Rmat}) by clockwise (anticlockwise)
$90^0$ rotations 
\begin{equation*}
  \widetilde{{\cal R}}_{\alpha\beta}^{\mu\nu}(v)=
  {{\cal R}}^{\beta\alpha}_{\mu\nu}(v) \qquad
  \overline{{\cal R}}_{\alpha\beta}^{\mu\nu}(v)=
  {{\cal R}}^{\alpha\beta}_{\nu\mu}(v).
\end{equation*}

\begin{figure}[tbp]
  \begin{center}
    \includegraphics[width=0.45\textwidth]{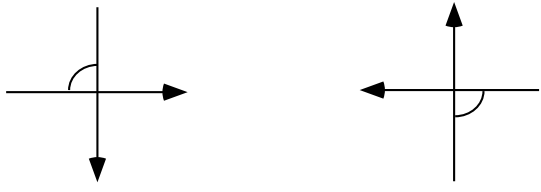}
    \caption{ In the same spirit to Fig.\ref{Rmatfig},
  R matrices,
	$ \widetilde{{\cal R}}$  and  $\overline{{\cal R}}$ are
	depicted as above.
	  }
	  \label{rmatsp}
 \end{center}
\end{figure}
The standard initial condition of
the $R^{\vee}$ matrix and (\ref{logderi}) lead to significant relations,
\begin{equation}
  {\cal T}(u)={\cal T}_R\, {\rm e}^{u{\cal H}/J+{\cal O}(u^2)}, \qquad
  {\overline{\cal T}}(u)={\cal T}_L\, {\rm e}^{u{\cal H}/J+{\cal O}(u^2)},
\label{qtm-exp}
\end{equation}
where ${\overline{\cal T}}$ is defined  in analogy to (\ref{transfer})
, replacing ${\cal R}$ by  $\overline{{\cal R}}$.
${\cal   T}_{R,L}$  are  the   right-  and left-shift   operators,
respectively and they commute with the Hamiltonian.

We are ready to apply Trotter formula;
by substitution 
\begin{equation}
u=-J \beta/N,
\label{u}
\end{equation}
we find 
\begin{equation}
  \Big({\cal T}(u)\,\overline{{\cal T}}(u)\Big)^{N/2}=
   {\rm e}^{-\beta{\cal H}+{\cal O}(1/N)}.
\label{qtm-element}
\end{equation}
where  $\beta$  denotes the inverse  temperature.
$N$ is a large integer ``Trotter'' number, interpreted as 
a fictitious system size in a virtual direction.
Thus the  partition function  of the quantum system (size $L$,
inverse temperature $\beta$) 
\begin{equation}
  Z=\lim_{N\to\infty}{\rm Tr}
  \Big({\cal T}(u)\,\overline{{\cal T}}(u)\Big)^{N/2},
\label{Z}
\end{equation}
is identical to the partition function of an inhomogeneous 2$S$+1
vertex model with alternating rows on
a virtual 2D lattice of size $L\times N$ .
\begin{figure}[tbp]
  \begin{center}
    \includegraphics[width=0.4\textwidth]{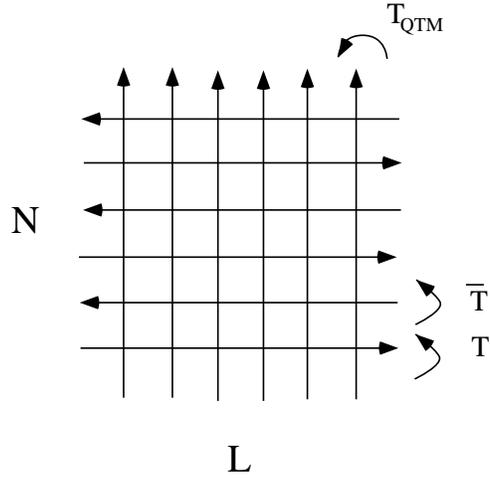}
    \caption{ 
	Graphic representation of the partition function on
	a virtual two dimensional lattice of $N\times L$.
	The operators ${\cal T}, \overline{{\cal T}}$ transfer states
	from bottom to
	top, while ${\tilde {\cal T}}(u)$ and 
	${\cal T}_{\rm QTM}$ do from right to left.
	  }
	  \label{zmnpic}
 \end{center}
\end{figure}
See (Fig .\ref{zmnpic}).
Although the above mapping is exact, the expression (\ref{Z}) is
not yet efficient.
The eigenvalues of 
${\cal T}(u)\,\overline{{\cal T}}(u)$  are 
almost degenerate. Hence it is still difficult task to evaluate
the trace.
The intriguing point in \cite{MSPRB} is to consider 
a transfer matrix ${\tilde {\cal T}}(u)$
propagating in the "horizontal" direction.

This novel operator acting on $N$ sites, has gaps in
the eigenvalues provided $T>0$.
Here we adopt further sophisticate approach developed in
\cite{KluRSOS}-\cite{JSE8}.
Explicitly, we define QTM by
\begin{equation}
  {\cal T}_{\rm QTM}(u,x)=\sum_\mu\prod_{i=1}^{N/2}
  {\cal R}_{\alpha_{2i-1}\,\beta_{2i-1}}^{\mu_{2i-1}\,\mu_{2i}}(u+ix)
  \,\widetilde{{\cal R}}_{\alpha_{2i}\,\beta_{2i}}
            ^{\mu_{2i}\,\mu_{2i+1}}(u-ix),
\label{qtm}  
\end{equation}
which reduces to above-mentioned ${\tilde {\cal T}}(u)$ by putting $x=0$.
\begin{figure}[tbp]
  \begin{center}
    \includegraphics[width=0.45\textwidth]{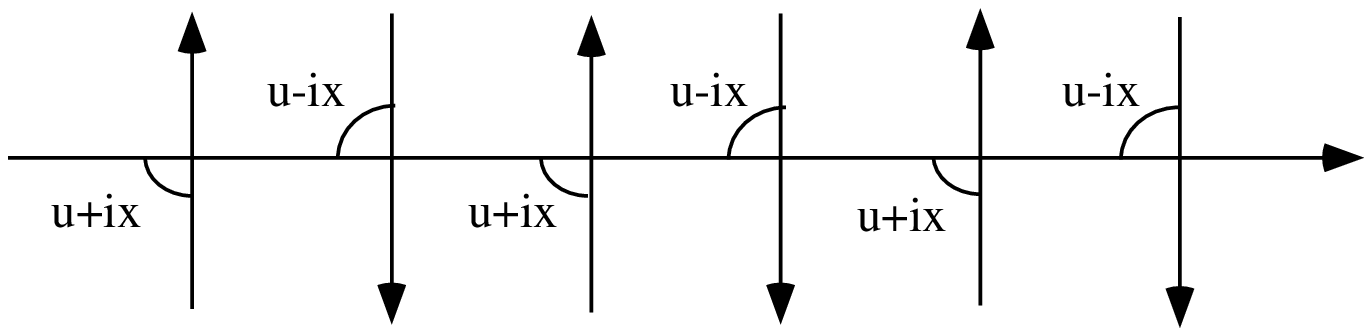}
    \caption{ 
	Graphic representation of  ${\cal T}_{\rm QTM}(u,v)$.
	  }
	  \label{qtmpic}
 \end{center}
\end{figure}
 Fig .\ref{qtmpic} represents QTM graphically .
Though the introduction of the extra parameter $x$ seems to
be redundant, there is a remarkable property; 
\begin{equation}
[{\cal T}_{\rm QTM}(u,x),{\cal T}_{\rm QTM}(u,x')]=0
\label{commute}
\end{equation}
by fixing $u$.
This originates from the fact that 
 ${\cal   R}$   and $\widetilde{{\cal R}}$
operators  possess   the   same intertwiner.
Thus for each $T$, one can associate a auxiliary complex plane $x$ to
the partition function.

Due to the gap in spectra, the free energy $f$
of 1D quantum spin chains is given only by the largest
eigenvalue  $\Lambda_{{\rm QTM}}(u,x)$,

\begin{equation}
  f=-\frac{1}{\beta}\lim_{L\to\infty}\frac{1}{L} \ln Z
   =-\frac{1}{\beta}\lim_{N\to\infty}
        \ln\Lambda_{{\rm QTM}}(u=\frac{-\beta J}{N},x=0).
\label{free-energy}
\end{equation}

This is the starting point of our analysis.
The difficulty in evaluating (\ref{free-energy})
lies in the $N$ dependence of the vertex weights.
The numerical extrapolation through finite $N$ studies 
may be plagued by
marginal perturbations\cite{KluZitt}.
The prescription is
to utilize the existence of complex plane $x$ for each $T$.
The analytic properties of ${\cal T}_{\rm QTM}$ and
suitably chosen auxiliary functions in the
$x-$ plane make the evaluation possible and transparent.

Before closing this section, we shall describe how to
modify the above relations in the presence of an
external magnetic field $H$, namely by inclusion
of the Zeeman term $-2 H \sum_i S_i^z $ to Hamiltonian (\ref{logderi}).
This contribution is described by diagonal operator $D(H)$,
\begin{equation*}
\left (
\begin{array}{c}
    {\rm e}^{-2 S\beta H} \\
    {\rm e}^{-(2S-1)\beta H} \\
    \vdots   \\
    {\rm e}^{2S\beta H} \\
\end{array}
\right )
\otimes
\left(
\begin{array}{c}
    {\rm e}^{-2S\beta H} \\
    {\rm e}^{-(2S-1)\beta H} \\
    \vdots   \\
    {\rm e}^{2S\beta H} \\
\end{array}
\right)
\otimes  \cdots
\otimes
\left(
\begin{array}{c}
    {\rm e}^{-2S\beta H} \\
    {\rm e}^{-(2S-1)\beta H} \\
    \vdots   \\
    {\rm e}^{2S\beta H} \\
\end{array}
\right).
\end{equation*}
Thus one only has to insert this inside the trace (\ref{Z}) 
\begin{equation*}
  Z=\lim_{N\to\infty}{\rm Tr}
  \Big({\cal T}(u)\,\overline{{\cal T}}(u)\Big)^{N/2} D(H).
\label{ZD}
\end{equation*}
In the rotated frame, the effect of the insertion of $D(H)$ is
translated to the boundary weight,
 $B(\mu_1)={\rm e}^{\beta \mu_1 H} $;
\begin{equation*}
  {\cal T}_{\rm QTM}(u,x)=\sum_\mu B(\mu_1) \prod_{i=1}^{N/2}
  {\cal R}_{\alpha_{2i-1}\,\beta_{2i-1}}^{\mu_{2i-1}\,\mu_{2i}}(u+ix)
  \,\widetilde{{\cal R}}_{\alpha_{2i}\,\beta_{2i}}
            ^{\mu_{2i}\,\mu_{2i+1}}(u-ix). 
\end{equation*}
%
%
\section{Fusion Hierarchy}
We consider a hierarchy of quantum transfer matrices acting on
$V_{k}^{\otimes_N}$.
Let $T_j(u,x)$ be a member of the hierarchy with
the auxiliary space  $V_j$. 

\begin{figure}[tbp]
  \begin{center}
    \includegraphics[width=0.45\textwidth]{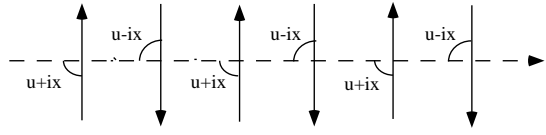}
    \caption{ 
     For $T_j(u,x)$, vertical arrows carry $V_k$ while 
	 horizontal (broken) arrow does $V_j$.
	 	  }
	 	  \label{qtmjp}
 \end{center}
\end{figure}

In other words,  it is the transfer matrix of the vertex model
of which spins $S(=k/2)$ are assigned to vertical edges 
and  spins $j/2$  to horizontal edges (Fig.\ref{qtmjp})
The quantity of our interest, $T_{\rm QTM}(u,x)$ coincides with
$T_k(u,x)$ apart from over-all normalization, which is specified later.

For brevity, we shall only give matrix elements of ${\cal R}(u)$ 
defining the most fundamental $T_1(u,x)$.
\begin{eqnarray*}
{\cal R}^{\pm 1/2,\pm 1/2}_{\ell,\ell} (u)&=&u+1\pm 2 \ell ,
\qquad
{\cal R}^{\ell'-\ell,\ell-\ell'}_{\ell',\ell} (u)= 
\sqrt{(k+2 +2{\rm min}(\ell,\ell')) 
 ( k -2min (\ell, \ell') ) }  \\
\end{eqnarray*}

where $\ell, \ell' \in \{-k/2 \cdots, k/2\} $ and
 $|\ell-\ell'| = 1$.
 Similar to (\ref{Rmat}), the corresponding  $R^{\vee}$  matrix has decomposition,
 \begin{equation}
  R^{\vee}(u) = (u+k+1) P_{k+1} +(u-k-1) P_{k-1}.
  \label{Rsing}
 \end{equation}
${\cal R}-$ matrix for $T_j(u,x)$  is obtained 
from the above elementary ${\cal R}(u)$  by $j-1$ times
fusion in the auxiliary space.
By the construction, arbitrary pairs 
in this hierarchy are commutative if
they share the same $u$, 
$$
[T_j(u,x),T_{j'}(u,x') ]=0.
$$
This is a generalization of  (\ref{commute}).
In the following, we fix $u$ for all QTMs and omit
the dependency on $u$.
Due to the consequential commutativity,
one needs not to distinguish operators
$T_j$  from their eigenvalues.

Then the explicit eigenvalue of the most elementary
transfer matrix $T_1(x)$ reads
\begin{eqnarray}
T_1(x) &=& 
 \phi_{+}(x-(k-1) i) \phi_{-}(x- (k+1)i)
  e^{\beta H} \frac{Q(x+2 i)}{Q(x)}  \np
 & & +\phi_{-}(x+ (k-1) i)\phi_{+} (x+(k+1)i) 
     e^{-\beta H} \frac{Q(x-2 i)}{Q(x)}   \np
  \phi_{\pm} (x) &:=& (x \pm i u) ^{N/2}, \np
 Q(x) &:=& \prod_{j=1}^{m} (x-x_j)  \np
\label{t1expr}
\end{eqnarray}

where $x_j, (j=1,\cdots, m)$ denotes the solution to
the Bethe ansatz equation:
$$
\frac{\phi_{-}(x_j+ (k-1) i)\phi_{+}(x_j+ (k+1) i)}
 {\phi_{+}(x_j- (k-1) i)\phi_{-}(x_j-(k+1) i)} = 
   - e^{2\beta H} \frac{Q(x_j+2 i)}{Q(x_j-2 i)}.
$$
The number of BAE roots, $m$, differs for different eigenstates generally,
and $m=Nk/2$ for the largest eigenvalue case.

By construction of the fusion hierarchy and 
from singularities of the intertwining operators (\ref{Rsing}) by tentatively
replacing $k \rightarrow j$,
 the following relation is valid:
\begin{eqnarray*}
T_j(x) T_1(x-i(j+1))&= &T_{j+1}(x-i) + g_j(x) T_{j-1}(x+i) \\
\qquad
g_j(x) &=& \phi_-(x-i(k+j+2)) \phi_-(x+i(k-j)) \phi_+(x-i(j+k)) \phi_+(x+i(k-j+2)).
\end{eqnarray*}
From this, one can prove the important  functional 
relation ($T-$ system) by induction \cite{KP},
\begin{eqnarray}
T_p(x+i) T_p(x-i) &=& f_p(x)+T_{p-1}(x) T_{p+1}(x), \quad(p \ge 1)  \np
 f_p(x) &:=& \prod_{j=1}^p \prod_{\sigma=\pm}
 \phi_{\sigma} (x+i\sigma(p-k-2j+1)) \phi_{\sigma} (x+i\sigma(k-p+2j+1)) \np
\label{Tsys}
\end{eqnarray}
and $T_0=1$.

By substituting (\ref{t1expr}) into (\ref{Tsys}),
we can successively
obtain  $T_p(x), (p \ge 2)$.
Explicitly, $T_p(x)$ consists of a sum of $p+1$ terms,
\begin{eqnarray}
T_p(x) &:=& \sum_{\ell=1}^{p+1} \lambda_{\ell}^{(p)}(x) , \np
\lambda_{\ell}^{(p)}(x) &:=& e^{\beta H(p+2-2\ell)}
\psi_{\ell}^{(p)}(x)
\frac{ Q(x+i(p+1)) Q(x-i(p+1))}{Q(x+i(2\ell-p-1)) Q(x+i(2\ell-p-3)) }, \np
\psi_{\ell}^{(p)}(x) &:=&
\prod_{j=1}^{p-\ell+1} \phi_-(x+i(p-k-2j)) \phi_+(x+i(p-k+2-2j)) \np
 & & \times \prod_{j=1}^{\ell-1} 
    \phi_-(x-i(p-k+2-2j)) \phi_+(x-i(p-k-2j)).  \np\label{tj}
\end{eqnarray}
As is previously noted, 
$T_{k}(x)$ has a normalization trivially different from
$T_{\rm QTM}(u,x)$ in the previous section:

\begin{eqnarray}
T_{\rm QTM}(u,x)&=&
  \frac{T_{k}(x)}{\prod_{p=1}^{k} \phi_0(2i p)} \np
\phi_0(x) &:=& x^{N/2} \np
\label{normalize}
\end{eqnarray}

In the original problem of the spin-$S$ chain,
only $T_{k}(x)$ is of our interest.
The auxiliary $T_j$'s, however, make the evaluation of  $T_{k}(x)$ 
transparent as is shown in the following.
%
\section{Auxiliary Functions and Functional Relations among them }

To explore the analyticity of the transfer matrix $T_{k}(x)$,
we introduce $k+1$ auxiliary functions.
The first $k-1$, functions,  $\{ y_j(x) \}$ are already
found in several literatures and have 
sound basis in the $sl_2$ fusion hierarchy.
They are defined by
\cite{KP, KNS}
$$
y_j(x) :=\frac{T_{j-1}(x) T_{j+1}(x) }{f_j(x)}, \qquad j\ge 1.
$$
The functional relations among them are sometimes referred to as
$Y-$ system:
\begin{eqnarray}
y_j(x+i) y_j(x-i) &= &Y_{j-1}(x)Y_{j+1}(x), \qquad j \ge 1 \np
Y_j(x) &:=& 1+y_j(x)
\label{Ysys}
\end{eqnarray}
and $y_0(x):=0$ which is a consequence of eq.(\ref{Tsys}).
Note that the $Y-$ system is not truncated to a finite set in this
case.  
The $k-1$-th equation, which characterizes $y_{k-1}(x)$
inevitably contains $y_{k}(x)$ in r.h.s., and so on.
Thus another device is needed to construct a finite set of auxiliary
functions satisfying a complete and closed set of functional relations.
The remaining two functions $\mathfrak{b}(x), 
\bar{\mathfrak{b}}(x)$ and their
"relatives" $\mathfrak{B}(x):=1+\mathfrak{b}(x), 
\bar{\mathfrak{B}}(x):=1+\bar{\mathfrak{b}}(x)$ play this role.
We define them by ratios of $\lambda$'s in $T_{k}(x)$ as, 
\begin{eqnarray}
\mathfrak{b}(x) &:=& 
\frac{\lambda_1^{(k)}(x+i)+\cdots+\lambda_k^{(k)}(x+i)}
         {\lambda_{k+1}^{(k)}(x+i)} \np
\bar{\mathfrak{b}}(x) &:=& 
\frac{\lambda_2^{(k)}(x-i)+\cdots+\lambda_{k+1}^{(k)}(x-i)}
    {\lambda_{1}^{(k)}(x-i)}. \np
\label{bdef}
\end{eqnarray}
The following relations are direct consequences of the above
definitions:
\begin{eqnarray}
\mathfrak{B}(x) \lambda_{k+1}^{(k)}(x+i) &=& 
     e^{-k\beta H} \mathfrak{B}(x) 
     \prod_{\sigma=\pm} \prod_{j=1}^k \phi_{\sigma} (x+(2j+\sigma)i)
          \frac{Q(x-ik)}{Q(x+ik)} = 
               T_{k}(x+i) \np
\bar{\mathfrak{B}}(x) \lambda_1^{(k)}(x-i) &=& 
     e^{k\beta H} \bar{\mathfrak{B}}(x) 
     \prod_{\sigma=\pm} \prod_{j=1}^k \phi_{\sigma} (x-(2j-\sigma)i)
          \frac{Q(x+ik)}{Q(x-ik)} 
         = T_{k}(x-i).
  \np
\label{BTrel}
\end{eqnarray}
We have $k-1$ equations for $y_j, (j=1,\cdots, k-1)$ in terms of
$Y_j(x), (j=1,\cdots, k-1), \mathfrak{B}(x)$ and 
$\bar{\mathfrak{B}}(x)$.
The first $k-2$ equations are chosen directly from the $Y-$ system.
In the $k-1$ th equation, we rewrite
$Y_{k}(x)$ in the rhs of $Y-$ system ($j=k-1$ in (\ref{Ysys}))
by $\mathfrak{B}(x) \bar{\mathfrak{B}}(x)$, thanks to (\ref{BTrel})
, the definitions of $y_k, Y_k$ and the functional relation (\ref{Tsys}):

\begin{equation}
y_{k-1}(x-i) y_{k-1}(x+i) = Y_{k-2}(x) 
            \mathfrak{B}(x) \bar{\mathfrak{B}}(x).
\label{YBrel}
\end{equation}

Finally, equations for $\mathfrak{b}$'s in terms of 
$Y_j(x), (j=1,\cdots, k-1), \mathfrak{B}(x)$ and 
$\bar{\mathfrak{B}}(x)$
are to be found.
By comparing explicit forms, one finds
\begin{eqnarray}
\mathfrak{b}(x) &=& e^{\beta(k+1) H}
   \prod_{\sigma=\pm} 
   \frac{\phi_{\sigma}(x+i\sigma )}
        {\prod_{j=1}^k \phi_{\sigma}(x+ (2j+\sigma)i) }
      \frac{Q(x+i(k+2))}{Q(x-ik)} T_{k-1}(x)   \np
\bar{\mathfrak{b}}(x) &=&e^{-\beta(k+1) H}
   \prod_{\sigma=\pm} 
   \frac{\phi_{\sigma}(x+i\sigma )}
        {\prod_{j=1}^k \phi_{\sigma}(x- (2j-\sigma)i) }
      \frac{Q(x-i(k+2))}{Q(x+ik)} T_{k-1}(x).  \np  
\label{bTrel}
\end{eqnarray}
Note that $T_{k-1}(x)$ is presented by $Y_{k-1}(x)$:
\begin{equation}
T_{k-1}(x-i) T_{k-1}(x+i) = f_{k-1}(x) Y_{k-1}(x)
\label{TYrel}
\end{equation}

which originates directly from definitions of $y_{k-1}, Y_{k-1}$ and
the functional relation (\ref{Tsys}).

In what follows, we analyze these functional relations via
the Fourier transformation.
One denotes $\widehat{dl}\mathfrak{b}[q]$ to 
mean the Fourier transformation
of the logarithmic derivative of $\mathfrak{b}(x)$,
$$
\widehat{dl}\mathfrak{b}[q]:=
\int_{-\infty}^{\infty} \frac{d \log \mathfrak{b}(x)}{dx} e^{iqx} dx,
$$
and similarly for other functions.
Under some assumptions on analytic properties of the
auxiliary functions, the above functional relations 
can be transformed into algebraic equations in the Fourier space.
Roughly speaking, one can solve $\widehat{dl}Q[q]$ functions 
in terms of $\widehat{dl}\mathfrak{B}[q]$ and 
$\widehat{dl}\bar{\mathfrak{B}}[q]$ by deleting 
$\widehat{dl}T_k[q]$ from algebraic
equations originated from
(\ref{BTrel}).
Similarly $\widehat{dl}T_{k-1}[q]$ is solved by 
$\widehat{dl}Y_{k-1}[q]$ from (\ref{TYrel}).
Substituting these results into Fourier transformations
of logarithmic derivatives of (\ref{bTrel}),
one finds expressions of $\widehat{dl}\mathfrak{b}[q],
 \widehat{dl}\bar{\mathfrak{b}}[q]$ 
in terms of
$\widehat{dl}\mathfrak{B}[q], \widehat{dl}\bar{\mathfrak{B}}[q]$ 
and $\widehat{dl}Y_{k-1}[q]$.
After the inverse Fourier transforming  and the 
integration over $x$, we shall find 
the desired finite set of equations.
We will make the above mentioned analytic assumptions explicit 
and examine  them in the next section.

Before going into details, let us discuss the physical
interpretation of the above functions.
As is argued in \cite{ReshetS}, the $S-$matrix
of excitations in the spin-$S$ model factorizes into
two pieces: spin-$\frac{1}{2}$ SU(2) $S-$matrix and RSOS $S-$matrix of 
$sl_2$ level $k=2S$ .
This is consistent with the general expectation that
the underlying field theory is the level-$k$ SU(2) WZWN model.
The latter is known to "decompose" into Gaussian and $Z_{k}$
parafermionic field theories \cite{FZ}.
One  finds, see for instance \cite{ABF}, 
several evidences for the
equivalence between the $sl_2$ RSOS model in regime II and the
$Z_{k}$ parafermion field theory in the scaling limit.
In the present description, 
$\mathfrak{b}, \bar{\mathfrak{b}}$ are to be identified 
with up- and down- spinons.
As we will see later, only they couple to the magnetic field
directly.
For the $S=1/2$ case, there are further direct evidences for
this identification \cite{JSch}.
On the other hand, $\{ y_j(x) \}$ are insensitive to the 
external field.
We are led to identify $y_j(x)$ in our choice as the RSOS piece
of the excitations. 
The RSOS model possesses a subset of the $Y-$ system (\ref{Ysys}).
The additional condition $y_{k}=0$ for the model leads to
the truncated set of equations among $k-1$
$y$'s \cite{BR, KP}.
In the present problem, at sufficiently low temperatures,
one observes,
$|\mathfrak{b}|, |\bar{\mathfrak{b}}| \sim 0$ for $x<< \ln \beta$. 
Thus the substitution of 
 $\mathfrak{B}=\bar{\mathfrak{B}}=1$ 
in (\ref{YBrel}) might be legitimate in the vicinity of
the origin.
Then the resultant approximated $Y-$ system coincides with that
of the RSOS model.
In this sense, (\ref{YBrel})  represents a gluing relation between 
spinon and RSOS parts of excitations.

%
%
\section{Nonlinear Integral Equations}

We derive the nonlinear integral equations among 
auxiliary functions introduced in the previous section.
The crucial observation is, that all nontrivial
zeros and singularities
of these functions are determined by zeros of
$Q(x)$ and $T_j(x), (j=1, \cdots, k)$.
For the largest eigenvalue sector of $T_k(x)$, 
zeros of $Q(x)$ form so-called $k-$ strings.
Imaginary parts of zeros are approximately
located at $(k+1)- 2\ell, \quad \ell=1,\cdots, k$.
For later use, we introduce notations,
\begin{equation}
\Psi_1(x):= Q(x-ik), \qquad \Psi_2(x):=Q(x+ik).
\label{defPsi}
\end{equation}
Empirically, similar patterns are found for zeros of $T_j(x)$:
they distribute approximately on lines,
$\Im x= \pm( k+j-2 \ell), \ell=0, \cdots, j-1$.
We assume that these observation from numerics
with fixed $N$ 
is valid and that the deviations 
from lines are very small in the limit $N \rightarrow \infty$. 
Then one deduces the
following ansatz on the strips where auxiliary functions
are Analytic, NonZero and  have Constant  asymptotic behavior (ANZC).

\begin{eqnarray*}
&&\mathfrak{b}(x), \mathfrak{B}(x) \qquad -1< \Im x \le 0, \\
&&\bar{\mathfrak{b}}(x), 
  \bar{\mathfrak{B}}(x)  \qquad  0 \le \Im x <1, \\
&&y_j(x), Y_j(x) (j=1, \cdots, k-1), \quad
T_p(x), (p=1, \cdots, k)  \qquad  -1 \le \Im x \le 1 \\
&&\Psi_1(x), \qquad \Im x<0  \\
&&\Psi_2(x), \qquad \Im x>0
\end{eqnarray*}
We find it convenient to shift the definition
of the arguments in $\mathfrak{b}, \mathfrak{B}, \bar{\mathfrak{b}}$ 
and $ \bar{\mathfrak{B}}$.
To avoid confusion, these new functions are denoted as
$\mathfrak{a}$ etc,
$$
\mathfrak{a}(x) :=\mathfrak{b}(x-i\gamma),
\qquad 
\bar{\mathfrak{a}}(x) :=\bar{\mathfrak{b}}(x+i\gamma),
$$
and similarly for capital functions.
Here $0<\gamma<1/2$ is an arbitrary but fixed parameter.
Note that this is equivalent to adopt small shifts in the 
definition of the integration contours for the Fourier transformation.
Due to the ANZC properties of $\mathfrak{b}, \bar{\mathfrak{b}}$,
such modifications are almost trivial in the Fourier space.

Having identified ANZC strips, we revisit eqs(\ref{BTrel}).
Consider the integral,
$$
\int_{\cal C} \frac{d}{d z} \log T_k(z) e^{i q z} dz,
$$
where ${\cal C}$ encircles the edges of "square" :
$[-\infty-i, \infty-i]\cup[\infty-i,\infty+i ]\cup
[\infty+i,-\infty+i ]\cup[-\infty+i, -\infty-i]$
in counterclockwise manner.
Due to the ANZC property of
$ \frac{d}{d z} \log T_k(z)$ inside ${\cal C}$,
the following equation is valid from
Cauchy's theorem:
$$
0= 
\int^{\infty}_{-\infty}  \frac{d}{d x} \log T_k(x-i) e^{iq (x-i)} dx
 -\int^{\infty}_{-\infty} \frac{d}{d x} \log T_k(x+i) e^{iq (x+i)} dx.
$$
One substitutes eqs(\ref{BTrel}), rewritten in terms of
$\Psi_{1,2}(x), \mathfrak{A}(x), \bar{\mathfrak{A}}(x)$, 
into the above equation
 and derives  identities among
$\widehat{dl}\Psi_{1,2}[q], \widehat{dl}\mathfrak{A}[q], 
\widehat{dl}\bar{\mathfrak{A}}[q]$,
\begin{eqnarray*}
\widehat{dl}\Psi_1[q<0]  &=& 0\\
\widehat{dl}\Psi_1[q>0]  &=& 
\frac{e^{(1-\gamma)q}}{2\cosh q} dl \bar{\mathfrak{A}}[q]
   - \frac{e^{-(1-\gamma)q}}{2\cosh q} dl\mathfrak{A}[q]   \\
    &+& \pi i N \frac{e^{-k q} \sinh kq \cosh(1+u)q}{\cosh q \sinh q}  \\
\widehat{dl}\Psi_2[q<0]  &=& 
-\frac{e^{(1-\gamma)q}}{2\cosh q} dl \bar{\mathfrak{A}}[q]
   + \frac{e^{-(1-\gamma)q}}{2\cosh q} dl\mathfrak{A}[q]   \\
    &-& \pi i N \frac{e^{k q} \sinh kq \cosh(1+u)q}{\cosh q \sinh q}  \\
\widehat{dl}\Psi_2[q>0]  &=& 0,\\
\end{eqnarray*}
and $\widehat{dl}\Psi_1[q=0]=-\widehat{dl}\Psi_2[q=0]=\pi N ki$.
Similarly, one can derive an identity for
$\widehat{dl}y_j[q]$'s and $\widehat{dl}Y_j[q]$'s from (\ref{Ysys}), and
$\widehat{dl}T_{k-1}[q]$ and 
$\widehat{dl}Y_{k-1}[q]$ from (\ref{TYrel}).
Substituting these relations into the
original definitions of 
$\mathfrak{a}$ and $ \bar{\mathfrak{a}}$,
we obtain $k+1$ algebraic relations in
Fourier space. 
(Remember $\Psi_{1,2}(x)$ are related to $Q(x)$ by (\ref{defPsi}).
After taking the inverse Fourier transformation and 
integrating over $x$ once, we arrive at the
$k+1$ coupled nonlinear integral equations:
\begin{equation}
\left(
  \begin{array}{c}
           \log y_1(x) \\
           \vdots      \\
           \log y_{k-1}(x) \\
           \log \mathfrak{a}(x) \\
           \log \bar{\mathfrak{a}}(x) \\
  \end{array}
\right )
= 
\left(
  \begin{array}{c}
          0 \\
           \vdots      \\
           0 \\
           \beta H+ d(u,x-i\gamma) \\
          - \beta H+ d(u,x+i\gamma) \\
  \end{array}
\right)
+
\cal {K} * 
\left(
  \begin{array}{c}
       \log Y_1(x) \\
           \vdots      \\
           \log Y_{k-1}(x) \\
           \log \mathfrak{A}(x) \\
           \log \bar{\mathfrak{A}}(x) \\
  \end{array}
\right )
\label{nlie}
\end{equation}

where $({\cal K}*g)_i$ denotes the matrix convolution,
 $ \sum_j \int {\cal K}_{i,j}(x-y) (g(y))_j dy$
and the "driving"  function $d(u,x)$ reads 
$$
d(u,x) = \frac{N}{2} \int \frac{\sinh uq}{q \cosh q} e^{-iqx} dq.
$$
The integration constants ($\pm\beta H$) are fixed by comparing 
asymptotic values ($|x| \rightarrow \infty$) of both sides.

Explicitly the kernel matrix is given by

\begin{equation}
{\cal K}(x) :=
\left(
  \begin{array}{cccccccc}
       0,& K(x),&    0,& \cdots,&   0,&   0, &           0,&           0   \\
    K(x),&    0,& K(x),& \cdots,&   0,&   0, &            &                \\
       0,& K(x),&      &        &   0,&   0, &      \vdots&     \vdots     \\
   \vdots&      &      &     &  \vdots& \vdots&            &                \\
       0,&      &\cdots&        &   0,& K(x),&          0,&            0  \\
       0,&      &      &        &K(x),&    0,& K(x+i\gamma),&K(x-i\gamma) \\
           0,&      &\cdots&        &  0,&K(x-i\gamma)&
                                                  F(x),& -F(x+2i(1-\gamma)) \\
           0,&      &\cdots&        &  0,&K(x+i\gamma)&
                                                -F(x-2i(1-\gamma)),& F(x) \\
  \end{array}
\right)
\label{ksym}
\end{equation}
where  
\begin{eqnarray*}
K(x) &:=& \frac{1}{4 \cosh\pi x/2}  \\
F(x) &:=& \frac{1}{2\pi} 
 \int_{-\infty}^{\infty} \frac{e^{-|q|-iqx}}{2 \cosh q} dq.  \\
\end{eqnarray*}
$T_k$, in terms of these auxiliary functions,
 can be derived similarly.
Technically, we find it convenient to introduce,
\begin{equation}
T_k^R(x) :=\frac{T_k(x)}{\prod_{p=1}^k 
    \phi_{e(p)}(x-2i(k+1-p)) \phi_{e(p+1)}(x+2i(k+1-p)) }
\label{renormT}
\end{equation}
and $e(p)=+ (-)$ for $p=$ even (odd).
Then the product of two equations in
 (\ref{BTrel}) leads to a simple algebraic relation:
\begin{equation}
T^R_k(x-i) T^R_k(x+i) = 
\begin{cases}
\mathfrak{B}(x) \bar{\mathfrak{B}}(x) \text{ for $k$ even} \\
\mathfrak{B}(x) \bar{\mathfrak{B}}(x) 
 \frac{\phi_-(x+i)\phi_+(x-i)}{\phi_-(x-i)\phi_+(x+i)}  \text{ for $k$ odd}\\
\end{cases}
\label{defTR}
\end{equation}
From ANZC property of the both sides of (\ref{defTR})
in appropriate strips,
the logarithmic derivative of them reduces to
a simple algebraic equation in Fourier space.
 Taking account of the normalization (\ref{normalize}) and (\ref{renormT})
, we have
\begin{eqnarray}
\log \Lambda_{QTM}(u,x) &=& \log \Lambda_{QTM}^{(0)}(u,x)  \np
& & +\int \frac{\log \mathfrak{A}(y)}{4 \cosh \pi/2(x-y+i\gamma)}dy+
   \int \frac{\log \bar{\mathfrak{A}}(y)}{4 \cosh \pi/2(x-y-i\gamma)} dy  \np
\log \Lambda_{QTM}^{(0)}(u,x)
 &:=&  -  \delta_{k=1 (mod 2)} 
           \frac{N}{2} \int e^{-iqx-|q|}\frac{\sinh uq}{q\cosh q} dq  
           \label{QTMA1}                          \\
 & & +\sum_{p=1}^k 
  \log\{  \frac{\phi_{e(p)}(x-2i(k+1-p)) \phi_{e(p+1)}(x+2i(k+1-p))}
       {\phi_0(2ip)^2 }   \}. \np
\label{QTMA2}
\end{eqnarray}

Finally put $u=-\beta J/N$ and 
send $N \rightarrow \infty$ {\it analytically}.
This merely amounts to replacements:
\begin{eqnarray}
d(u,x) &\rightarrow&  D(x) = -\frac{\beta J \pi}{ 2 \cosh \pi/2 x} \np
\log \Lambda_{QTM}^{(0)}(u,x=0) &\rightarrow&
        -\beta e_0 = -\frac{\beta J}{2} \sum_{j=1}^k \frac{(-1)^{k-j}}{j} +
                \delta_{k=1 (mod 2)} \beta J \log 2.
\label{Ninf}
\end{eqnarray}
Note that $e_0$ coincides with the known ground state energy 
\cite{Bab} after a trivial shift which stems from the
difference in the normalization of $R^{\vee}$.
The $k+1$ coupled nonlinear integral equations and 
$\log \Lambda_{QTM}$
do not carry the fictitious parameter $N$ any longer.
They efficiently describe the thermodynamics of the 
solvable spin-$S$ XXX model.
For an illustration, the specific heat for $S=1/2,1,3/2$ is evaluated
for a wide range of temperature and plotted below.
\begin{figure}[tbp]
  \begin{center}
    \includegraphics[width=0.45\textwidth]{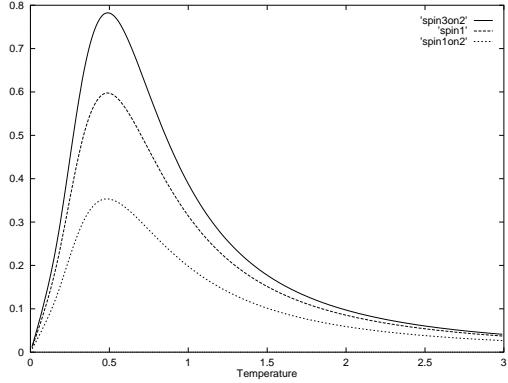}
    \caption{specific heats for $S=$ 3/2, 1 and 1/2 from top to bottom}
 \end{center}
\end{figure}
Each curve is produced by 10-30 minutes CPU time calculation on a 
Micro Sparc Work Station.
In the next section, we derive the low temperature properties
using (\ref{nlie}), (\ref{QTMA1}) and (\ref{Ninf}).
%
%
\section{Analytic Evaluation of the Low Temperature Asymptotics}

We consider $T\rightarrow 0$ for the vanishing magnetic field.
In sufficiently low temperature regime, 
$\mathfrak{a}, \log{\mathfrak{A}}
$ shows a sharp crossover behavior like
a step function:
$|\mathfrak{a}|, |\log{\mathfrak{A}}| \ll 1$ 
for $|x| < \frac{2}{\pi} \log \pi \beta J $
and 
 $|\mathfrak{a}|,|\log{\mathfrak{A}}| =O(1)$
for $|x| > \frac{2}{\pi} \log \pi \beta J $.
Thus the following scaling functions \cite{KBP} control the
asymptotic behavior:
\begin{eqnarray}
la^{\pm}(\xi) &:=& 
\log \mathfrak{a} (\pm \frac{2}{\pi}(\xi+\log \pi \beta J)),
\qquad 
lA^{\pm}(\xi):= 
\log \mathfrak{A} (\pm \frac{2}{\pi}(\xi+\log \pi \beta J)),
\np
l\bar{a}^{\pm}(\xi) &:=& 
\log \bar{\mathfrak{a}} (\pm \frac{2}{\pi}(\xi+\log \pi \beta J)),
\qquad 
l\bar{A}^{\pm}(\xi):=
 \log \bar{\mathfrak{A}} (\pm \frac{2}{\pi}(\xi+\log \pi \beta J)),
\np
ly_p^{\pm}(\xi) &:=& 
\log y_p (\pm \frac{2}{\pi}(\xi+\log \pi \beta J)),
\quad
lY_p^{\pm}(\xi) := 
\log Y_p (\pm \frac{2}{\pi}(\xi+\log \pi \beta J)).
\end{eqnarray}

In terms of these scaling functions, NLIE are
expressed by,
\begin{eqnarray}
\left(
  \begin{array}{c}
          ly_1^{\pm}(\xi) \\
           \vdots      \\
           ly_{k-1}^{\pm} (\xi) \\
           la^{\pm}(\xi)  \\
           l\bar{a}^{\pm} (\xi) \\
  \end{array}
\right ) 
&= &
\left(
  \begin{array}{c}
           0 \\
           \vdots      \\
           0 \\
          - e^{-\xi \pm i\gamma \pi/2} \\
          - e^{-\xi \mp i\gamma \pi/2} \\
  \end{array}
\right)
+
\bar{\cal K} * 
\left(
  \begin{array}{c}
           l Y_1^{\pm}(\xi) \\
           \vdots      \\
           l Y_{k-1}^{\pm}(\xi) \\
           lA^{\pm}(\xi) \\
           l{\bar{A}}^{\pm}(\xi) \\
  \end{array}
\right )  \np
\bar{\cal K}(\xi) &=& \frac{2}{\pi} {\cal K}(\frac{2 \xi}{\pi}) \np
\label{scalingnlie}
\end{eqnarray}
Note that neglect of small corrections $\sim O(T)$ leads
to the decoupling equations for $\pm$.

The thermal contribution (the second and the third term
in (\ref{QTMA1})) to 
$\lim_{N \rightarrow \infty} 
\log \Lambda_{\rm QTM}(u=-\frac{\beta J}{N},x)$
reads,

\begin{eqnarray}
&&\frac{e^{\pi x /2 }}{\pi^2 \beta J} 
 [ e^{i \gamma \pi/2 } \int e^{-\xi} lA^{+} d \xi + 
    e^{-i \gamma \pi/2 } \int e^{-\xi} l\bar{A}^{+} d \xi ] \np
&& +
\frac{e^{-\pi x/2 }}{\pi^2 \beta J} 
 [ e^{-i \gamma \pi/2 } \int e^{-\xi} lA^{-} d \xi + 
    e^{i \gamma \pi/2 } \int e^{-\xi} l\bar{A}^{-} d \xi ]. \np
\label{T0free}
\end{eqnarray}
The crucial observation in \cite{KBP} is that 
one needs not solve (\ref{scalingnlie}) to evaluate (\ref{T0free})
provided that the kernel matrix function satisfies a symmetry, 
$ {\cal K}_{i,j}(x-y)={\cal K}_{j,i}(y-x) $.
This property is valid in the present case. See (\ref{ksym}).

We define $F_{\pm}$ by
\begin{eqnarray}
F_{\pm}&:=& \int_{-\infty}^{\infty}
  \sum_{p=1}^{k-1} [ (\frac{d}{d\xi} ly_{p}^{\pm}) lY_{p}^{\pm}
  - (\frac{d}{d\xi} l Y_{p}^{\pm}) ly_{p}^{\pm} ] d\xi \np
& & +
\int_{-\infty}^{\infty}[
 (\frac{d }{d\xi} l a^{\pm})lA^{\pm} +
  (\frac{d }{d\xi}l\bar{ a}^{\pm})l\bar{A}^{\pm}  -
 (\frac{d }{d\xi}l A^{\pm}) la^{\pm} -
 \frac{d}{d\xi} l\bar{ A}^{\pm})l\bar{a}^{\pm}] d\xi.
\label{Fdef}
\end{eqnarray}
Then the trick in \cite{KBP} is as follows.
First, take the derivative of both sides of (\ref{scalingnlie}) and
multiply them by  a row vector
\begin{equation}
\Bigl( lY_1^{\pm}(\xi), \cdots, lY_{k-1}^{\pm}(\xi), lA^{\pm}(\xi),
l\bar{A}^{\pm}(\xi)  \Bigr ).
\label{rvec}
\end{equation}
We call the resultant equality (A).
Second, multiply both sides of (\ref{scalingnlie}) by the derivative of the
row vector (\ref{rvec}), which is referred to as (B).
Finally, subtract both sides of (A) and (B), and integrate over $\xi$.
Then the lhs of the resultant equality is nothing but $F_{\pm}$.
Remarkably,  the most complicated terms in the rhs, like 
$$
 \int d\xi d \xi'  lY^{+}_i(\xi) 
 \frac{d \bar{{\cal K}}_{i,j}(\xi-\xi')}{d\xi}lY^{+}_j(\xi') 
$$
and 
$$
- \int d\xi d \xi'  \frac{d}{d\xi} lY^+_j(\xi)
    \bar{{\cal K}}_{j,i}(\xi-\xi') lY^+_i(\xi')
$$
cancel with each other.
To be precise, the first integral can be converted step by step,
\begin{eqnarray*}
&=&-\int d\xi d \xi'  lY^+_i(\xi) 
    \frac{d \bar{{\cal K}}_{i,j}(\xi-\xi')}{d\xi'}lY^+_j(\xi')   \\
&=& -\int d\xi d \xi'  lY^+_i(\xi) 
    \frac{d \bar{{\cal K}}_{j,i}(\xi'-\xi)}{d\xi'}lY^+_j(\xi')   \\
&=& \int d\xi d \xi'  lY^+_i(\xi) 
    \bar{{\cal K}}_{j,i}(\xi'-\xi) \frac{d}{d\xi'} lY^+_j(\xi')  \\
&=& \int d\xi d \xi'   \frac{d }{d\xi} lY^+_j(\xi) 
    \bar{{\cal K}}_{j,i}(\xi-\xi')  lY^+_i(\xi')  \\
\end{eqnarray*}
where the symmetry of the kernel matrix, partial integration and
the change of integration variables $\xi \leftrightarrow \xi'$ are used.

Similar cancellation happens for other terms. 
and the following equality results,
\begin{equation}
F_{\pm} = 2 \int [ e^{-\xi\pm i \gamma \pi/2} lA^{\pm}  +
                   e^{-\xi\mp i \gamma \pi/2} l\bar{A}^{\pm} ] d\xi.
\end{equation}
The first thermal correction (\ref{T0free}) is thus given by
\begin{equation}
\frac{ e^{\pi x/2 } F_+}{2 \pi^2 \beta J} +
   \frac{ e^{-\pi x/2 } F_-}{2 \pi^2 \beta J}.
\end{equation}
To evaluate $F_{\pm}$ explicitly, one rewrites the
integration variable from
$\xi$ to $a, \bar{a}, y_p$.
For example, the first summation term in $F_{\pm}$ is transformed to
$$
\sum_{p=1}^{k-1} \int_{y_p^\pm (-\infty)}^{y_p^\pm (\infty)}
 dy ( \frac{ \log(1+y)}{y}-\frac{\log y}{1+y} )
=2\sum_{p=1}^{k-1} \{ L_+(y_p^\pm (\infty))- L_+(y_p^\pm (-\infty)) \}, 
$$
and similarly for others.
$L_+(x)$ is a dilogarithm function and is related to
Rogers' dilogarithm function $L(x)$ by  $ L_+(x) = L(x/(1+x))$,

\begin{eqnarray}
L_+(x) &:=& \frac{1}{2}\int_0^x 
  ( \frac{ \log(1+y)}{y}-\frac{\log y}{1+y} ) dy,\np
L(x) &:=& -\frac{1}{2} \int_0^x  (\frac{\log(1-y)}{y}+\frac{\log y}{1-y}) dy \np
\label{defRogers}
\end{eqnarray}

The asymptotic values of scaling functions are easily extracted.
For $x \rightarrow  \infty$, $a^{\pm}$ coincides with 
original $\mathfrak{b}$. 
Thus one derives the limiting value
by its definition (\ref{bdef}) in terms of $\lambda^{(k)}_p$'s.
Similar for $y^{\pm}_p(\infty)$.
For $x \rightarrow  -\infty$, one should rather
 consult (\ref{scalingnlie}).
We send the argument $x \rightarrow - \infty$ in both side
and solve the resultant algebraic equations.
The results are summarized as,
\begin{eqnarray}
a^{\pm}(- \infty)&=&\bar{a}^{\pm}(- \infty)=0, 
\qquad 
a^{\pm}( \infty)=\bar{a}^{\pm}( \infty)=k, \np
y_p^{\pm} (-\infty) &=& 
\frac{\sin\frac{\pi p}{k+2} 
\sin\frac{\pi (p+2)}{k+2} }
         { \sin^2\frac{\pi }{k+2} }  \quad 1\le p \le k-1, \np
y_p^{\pm} (\infty) &=& p(p+2)  \quad 1\le p \le k-1.                         
\label{asym}
\end{eqnarray}
With these pieces of information, $F_{\pm}$ is now given by
$$
F_{+}=F_-= 2\sum_{p=1}^{k-1} 
    [ L(\frac{p(p+2)}{(p+1)^2}) - 
          L(   \frac{\sin\frac{\pi p}{k+2} \sin\frac{\pi (p+2)}{k+2} }
             { \sin^2\frac{\pi(p+1) }{k+2} }  )] +
        4L(\frac{k}{1+k})
$$
Finally we use three relations\cite{Kir}:

\begin{eqnarray*}
L(1) &=& L(x)+L(1-x) = \frac{\pi^2}{6} , \quad x \in [0,1]   \np
2 L(1) &=& 2 L(\frac{1}{n+1}) + 
    \sum_{j=0}^{n-1} L(\frac{1}{(1+j)^2}),\quad   n \in Z_{\ge 0}, \np
L(1) \frac{3 n}{n+2} &=& \sum_{j=0}^{n-1} 
L(  \frac{\sin^2\frac{\pi }{n+2}  }
         { \sin^2\frac{\pi(j+1) }{n+2} } ), \quad n \in Z_{\ge 0}
\end{eqnarray*}

which yield the neat result
$$
F_{\pm} = \frac{ \pi^2 k}{k+2}.
$$
Thus we conclude the low temperature asymptotics of free energy
\begin{equation}
f \sim e_0 - \frac{\pi}{6 v_s \beta^2} c(k),
\qquad c(k)= 3k/(k+2),
\label{centralc} 
\end{equation}
where $v_s= J \pi/2$ coincides with the known spin velocity\cite{Sogo} and 
$c(k)$ is nothing but the central charge of 
level-$k$ SU(2) WZWN model.
This is the desired result from the WZWN description
 of massless  quantum spin chains.
We remark that the final part of the calculation is
quite parallel to that in \cite{BabTsv} utilizing the same
 dilogarithm function identities.
There the spin-$S$
XXZ model  
is discussed via the standard string approach at "root of unity"
where  the number of strings is truncated finitely from the 
beginning.
%
%
\section{Spinon Characters}

The character formulae obviously depend on the base of the space.
Recently, the quasi-particle representation has been attracted much attention
in the context of long-range interacting model \cite{BLS,BPS,NakYam}, 
spectral decomposition of path space in lattice models  \cite{Nagoya},
or  in the 
statistical interacting picture of Bethe Ansatz solvable models
\cite{SUNY931}-\cite{BerkMC96}.
See also \cite{Fring, Gaite} in the different view points.
For the spin-1/2 case, it has been shown recently that the novel
thermodynamics formulation yields a natural spinon character 
and such character formula 
is  generalized to $\widehat{\mathfrak{sl}}(n)_{k=1}$
\cite{JSch}.
Thus it is tempting to find analogues for
$\widehat{\mathfrak{sl}}(n)_{k=2S}$.
The results given in the previous sections provide
the first step for the simplest $n=2$ case as discussed below.
Remark that we  consider "chiral-half", 
such that the only $+$ part contributions (say $F_+$ ) 
in the previous section
are taken into account.

The character needs the description of all excited states.
In the present contexts,  this information might be encoded in
the additional zeros of $\mathfrak{a}, \bar{\mathfrak{a} }, y$
and their capitals in their "physical strips".
Indeed, some low excitations are identified in such a way
\cite{JKSFusion,  KSS}, 
and corresponding excited state TBA are derived.
Such an analysis is of considerable interest, however
it requires extensive numerical efforts.
We leave it as a future problem and make a short-cut detour here
employing the  strategy in \cite{KNSch}.

The central charge is described by the dilogarithm function
of which the integration contour is simple.
On the other hand, one can define an analytically continued dilogarithm function
$L_{\cal C}(z)$.
This is established by adopting general contour ${\cal C}$ 
for the integration contour of the dilogarithm function.
We then generalize a successful observation from specific examples;
 all excitation spectra, or effective central charge $c_{\rm eff}$, 
in the conformal limit shall be described by $ L_{\cal C}(z)$. 
Namely, the replacement of  a simple contour in the integral representation of
the dilogarithm function by complex one leads to an excited state.
Regarding $c_{\rm eff}$ as a function of  $L_{\cal C}(z)$,
the summation of $q^{-c_{\rm eff}/24}$ over a certain set of
${\cal C}$ is thus expected to reproduce affine characters.
(Readers should not confuse this formal variable $q$,
the standard notation in this field,  with the Fourier
variable used in  previous sections.)

Let us be more precise.
By ${\cal C}$ we denote a contour starting from $f_-$ and 
terminating at $f_+$, such that it 
crosses firstly $[1, \infty)$ $\eta_1 (\ne 0)$ times  then 
crosses  $(-\infty,0]$\ $\xi_1$ times  then
again $\eta_2$ times w.r.t.  $[1, \infty)$ and so on.
The intersections are counted as $+1 (-1) $
  if the contour goes across the cut
 $[1, \infty)$  in the counterclockwise (clockwise) manner and 
$(-\infty,0]$   in the clockwise (counterclockwise) manner.
(Note the definition is slightly different
 from \cite{KNSch, JSch}.)
We denote this by 
${\cal C}[f_-,f_+|\{\xi_1, \xi_2, \cdots \}|\{\eta_1, \eta_2, \cdots\}]$.
\begin{figure}[tb]
  \begin{center}
    \includegraphics[width=0.45\textwidth]{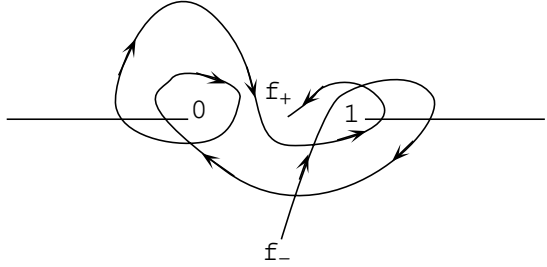} 
    \caption{A contour ${\cal C}[f_-,f_+|\{2 \}|\{-1,1\}]$.} 
    \label{fig:figcv}
  \end{center}
\end{figure}
The set of contours are  parameterized by ${\cal S}
=\{ {\cal C}[f_-^{(p)},f_+^{(p)}|\xi^{(p)}_1, \cdots, |
        \eta^{(p)}_1, \cdots,   ] \}$, 
where
$f^{(p)}_{\pm} = y^+_p(\pm \infty)/Y^+_p(\pm \infty), (1\le p\le k-1)$, 
$f^{(k)}_+=\mathfrak{a}^+(\infty)/\mathfrak{A}^+(\infty)$, 
$f^{(k+1)}_+=\bar{\mathfrak{a}}^+(\infty)/\bar{\mathfrak{A}}^+(\infty)$
and $f^{(k)}_-=f^{(k+1)}_-=0$.

In the absence of additional zeros of auxiliary functions
in the "physical strip",  we have 
 \begin{equation}
  0=\log(f_+^{(p)})-
    \sum_{p'} \mathfrak{g}_{p,p'}  \log(1-f_+^{(p')}),
 \label{asytba}
\end{equation} 
 where the ``statistical interaction'' matrix $\mathfrak{g}$ is related 
to the zero mode of the Fourier transformation of
the  kernel matrix  (\ref{ksym}) by
\begin{equation}
\mathfrak{g} =I - {\cal K}[q=0].
\label{defg}
\end{equation} 
This is the situation we have treated in previous sections,
and (\ref{asytba}) follows from (\ref{scalingnlie}).
One replaces
$\log$  in (\ref{asytba})  by  an analytically 
continued one, ${\rm Log}_{\cal  C} (z)$  in excited states:
\begin{equation}
 \pi D^{(p)}= {\rm Log}_{{\cal  C}^{(p)}}(f_+^{(p)})-
    \sum_{p'} \mathfrak{g}_{p,p'}{\rm Log }_{{\cal  C}^{(p')}}(1-f_+^{(p')}),
\label{excTBA}
\end{equation}
where ${\rm Log}_{{\cal  C}^{(p)}}(f_+^{(p)})=
\log (f_+^{(p)})-2\pi i \sum_{\ell} \xi^{(p)}_{\ell}$
and so on.
Here $D^{(p)}$ is introduced for consistency of both sides, and is
  interpreted as "chemical potential"  \cite{KNSch}. 
On the other hand, $D^{(p)}$ should originate 
from the zeros of auxiliary functions in the
physical strips.
One may be able to prove that  such $D^{(p)}$ actually
agree with one in (\ref{excTBA}), in principle.
Though such microscopic derivation is 
yet to be done, we assume the coincidence in the following.
 
Now, the excitation spectrum is solely implemented in the
effective central charge $c_{\rm eff}({\cal S})$
\begin{eqnarray}
  c_{\rm eff}({\cal S}) =
  \frac{6}{\pi^2} \sum_p 
   \Bigl (  L_{{\cal C}^{(p)}}(f_-^{(p)}, f_+^{(p)}) 
            -\frac{\pi i}{2} D^{(p)} 
             {\rm Log}_{{\cal C}^{(p)}}(1-f_+^{(p)})   \Bigr ).
	    \label{ceff0}
 \end{eqnarray}
Here $D^{(p)}$ term is included by hand,
so as to match its interpretation as chemical potential \cite{KNSch}.
$L_{\cal C}$ is given by,
\begin{equation}
L_{\cal C}(f_-,f_+) =
-\frac{1}{2} \int_{\cal C}
  ( \frac{{\rm Log}_{\cal C}(1-z)}{z}+ \frac{{\rm Log}_{\cal C}(z)}{1-z}) dz.
\end{equation}
for a path ${\cal C}$ from $f_-$ to $f_+$. 

 After straightforward manipulations, one finds \cite{KNSch},
\begin{eqnarray}
 L_{{\cal C}^{(p)}}(f_-^{(p)}, f_+^{(p)})& =&
 L(f_+^{(p)})-L(f_-^{(p)})
 -\pi i \sum_{\ell} \xi^{(p)}_{\ell} \log(1-f_+^{(p)})-
 \pi i \sum_{\ell} \eta^{(p)}_{\ell} \log(f_+^{(p)})    \np
&& + 2 \pi^2 (\sum_{\ell} \xi^{(p)}_{\ell}) (\sum_{\ell} \eta^{(p)}_{\ell} )
  -4\pi^2 \sum_{\ell}  \xi^{(a)}_{\ell} 
                        (\eta^{(a)}_1+\cdots +\eta^{(a)}_{\ell}) .\np
\label{Lexpand}					
\end{eqnarray}
Remember $L(z)$ is defined in (\ref{defRogers}).
The substitution of (\ref{Lexpand}) in (\ref{ceff0}), using
(\ref{excTBA}) and the definitions of ${\rm Log}_{{\cal C}^{(p)}}$
leads to a remarkable result:
 $c_{\rm eff}$
  can only be written in terms of $c(k)$ (see (\ref{centralc})),
$\{ \xi_{\ell}^{(p)} \} $ and  $\{ \eta_{\ell}^{(p)} \} $,
 \begin{eqnarray}
 c_{\rm eff}({\cal S}) &=&  c(k) - 24 {\cal T}({\cal S}) ,  \qquad
  {\cal T}({\cal S}) =
  \frac{1}{2} \phantom{}^t \mathbf{n} \mathfrak{g}\mathbf{n} +
          \sum_{a=1}^{k+1} \sum_{\ell \ge 1} \xi^{(a)}_{\ell} 
                        (\eta^{(a)}_1+\cdots +\eta^{(a)}_{\ell})  \np
\phantom{}^t \mathbf{n} &=&
 (n^{(1)}, \cdots, n^{(k+1)})
\label{ceff}
\end{eqnarray}
where $n^{(p)}= \sum_{\ell} \eta^{(p)}_{\ell}$ .
Note that the explicit forms of $D^{(p)}$ 
 are not needed in the above transformation.

In the following, we shall argue that the summation of $q^{-c_{\rm eff}}$
over some subset  ${\cal O}$ of all possible contours
reproduce the character ${\rm ch}_j(z,q)$ 
of level $k$ WZWN model with spin-$j$
($j=$ some fixed  integer or half-integer).

We present necessary conditions for  such  ${\cal O}$ below.

\begin{itemize}
\item   $\eta's$ and $\xi's$ are non-negative.
\end{itemize}
Such path can be parameterized by 
$$
{\cal C}[f^{(p)}_-, f^{(p)}_+ | \xi_1^{(p)}, \cdots, \xi_{n^{(p)}}^{(p)}
 | \underbrace{1,\cdots, 1}_{n^{(p)}}]
$$
and $\xi_{\ell}^{(p)}\ge 0, (p=1,\cdots,n^{(p)})$.
\begin{itemize}
\item For $p=k, k+1$,  
we require $\frac{n^{(k)}+n^{(k-1)}}{2}-j \in {\bf Z}_{\ge 0}$ in addition.
\item For $p \le k-1$, $\xi_{n^{(p)}}^{(p)}\ge 1$.
\end{itemize}
Graphically, one can associate a Young diagram $YD^{(p)}$
to a set of winding numbers: $\{\xi^{(p)}_{\ell} \},
\{\eta^{(p)}_{\ell} \}$
(or a Young diagram with tail for some cases in $p=k,k+1$).
First draw a line of length 1 downwards.
Next draw a line of length $\xi^{(p)}_1$ to the left.
Then again draw a line of length 1 downwards.
We continue this procedure $n^{(p)}$ times.
Finally draw a horizontal line from the starting point to the left
and also draw a vertical line from the end point upwards.
\begin{figure}[tb]
  \begin{center}
    \includegraphics[width=0.45\textwidth]{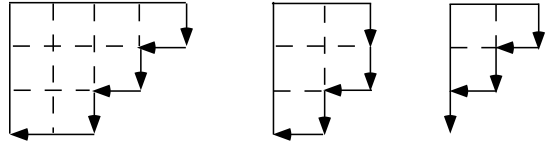} 
    \caption{Young diagrams $YD^{(p)}$ of $n^{(p)}=3$
          corresponding
	 to a set of winding numbers: $\{\xi \}=\{1,1,1 \},
	 \{ \eta \}=\{1,1,2\}$ (left), 
	 $\{\xi \}=\{1,1,1 \}, \{ \eta \}=\{0,1,1\}$ (middle)
	  and 
	  $\{\xi \}=\{1,1,1 \}, \{ \eta \}=\{1,1,0\}$ (right).
	 Arrows are put for a guide to eyes.
	 The most right diagram is termed "with tail" in the text. } 
    \label{fig:fig2ab}
  \end{center}
\end{figure}
See the Figure \ref{fig:fig2ab}.
Obviously, the number of boxes in the diagram is equal to 
the second term in ${\cal T}({\cal S})$.

 We allow for contours which are isomorphic to 
 a set of Young diagrams $YD^{(p)}, (p=1, \cdots k-1)$  such that
\begin{itemize}
\item
$n^{(k-2p+1)}=$ odd, $n^{(k-2p)}=$ even.
\item
By the definition, the depth of $YD^{(p)}$  is  $n^{(p)}$.
The width is restricted by the maximum value 
$w^{(p)}_{\rm max} $ which is 
determined by depths of "adjacent" diagrams:
$$
w^{(p)}_{\rm max}=
1/2(n^{(p-1)}+n^{(p+1)}-2 n^{(p)} +\delta_{2j,k-p}),\qquad(p=1,\cdots,k-2)
$$ 
for the fixed
$j$ and $n^{(0)}=0$ .
The case $p=k-1$ is exceptional,
$w^{(k-1)}_{\rm max} = 1/2(n^{(k-2)}+n^{(k)}+n^{(k+1)}-2n^{(k-1)}+
\delta_{j,1/2})$.
\end{itemize}

Under the above restrictions on ${\cal O}$, we find
\begin{eqnarray*}
q^{-\triangle(j) +j/2} {\rm ch}_j(z,q) &=& 
 \sum_{ {\cal S} \in {\cal O} } 
     q^{-c_{\rm eff}({\cal S})/24} z^{(n^{(k)} - n^{(k+1)})/2}  \np  
&=&
\sum_{n^{(k)}, n^{(k+1)} \ge 0} q^{-(n^{(k)}+n^{(k+1)})^2/ 4} 
 \Psi_{A_k}^{n^{(k)}+n^{(k+1)}}(u_j;q) {\cal S}_{n^{(k)}, n^{(k+1)}}(z;q).
\end{eqnarray*}
${\cal S}_{M,N}(z;q)$ stands for  contributions from 
$(M,N)$ spinons;
$$
{\cal S}_{M,N}(z;q) =\frac{1}{(q)_M (q)_N} z^{(M-N)/2}
$$
resulting from summations over "nodes" $\xi^{(p)}_{\ell}\ge 0,  \ell=1, \cdots, n^{(p)}$
and $p=k,k+1$.
$A_k$ is the Cartan matrix for 
$sl_{k+1}$ and  $\Psi_{A_k}$  denotes

$$
\Psi_{A_k}^{m_1} (u_j;q) = 
\sum_{m_2, m_3, \cdots, m_k} q^{1/4 m. A_k m}
 \prod_{i=2}^{k} 
 \begin{bmatrix}
  \frac{1}{2}((2-A_k).m +u_j)_i \\
  m_i \\
 \end{bmatrix}
$$

and $(u_j)_i= \delta_{i,2j+1}$.
The summations are taken 
over odd (even) positive integers for $m_{\rm even (odd)}$.
Note we redefine $n^{(k-\ell)}=m_{\ell+1}, \,\, 1 \le \ell \le k-1$.
The appearance of the Gaussian q-binomial
usually originates from combinatorics on the
truncated Bratteli diagram and is the reminiscence of
the RSOS model.
Here the origin is also simple. 
It comes from the restriction on the
width of Young diagrams.
We denote the number of boxes in a Young diagram $YD^{(p)}$ by $b^{(p)}$.
For fixed $\{ n^{(a)} \}$, ${\cal T(S) }$ assumes the same value
for diagrams having identical $b^{(p)}$.  
This multiplicity is given by
$p(n^{(p)}, w^{(p)}_{\rm max},  b^{(p)}) =$ the number of
partition of $b^{(p)}$ into at most $ w^{(p)}_{\rm max}$ part,
each $\le n^{(p)}$.
Thanks to the generating relation,
$$
 \sum_{b^{(p)}} p(n^{(p)},w^{(p)}_{\rm max},b^{(p)}) q^{b^{(p)}}
  = \begin{bmatrix}
              n^{(p)}+w^{(p)}_{\rm max}\\
							 n^{(p)}
		\end{bmatrix}
 =\begin{bmatrix}
    (n^{(p-1)}+n^{(p+1)}+\delta_{p,k-2j})/2\\
     n^{(p)}
   \end{bmatrix}
$$
we obtain the product of Gaussian q-binomial equivalent to
the one in $\Psi_{A_k}^{m_1} (u;q) $.

Our result implies that the Hilbert space of level $k$ SU(2) WZWN
model is isomorphic to a part of homotopy space of
Rogers' dilogarithm function.
We believe that this provides a novel interesting view point in the
prototype of CFT and deserves further examinations.
%
%
\section{Summary and Discussion}
In this report,  we formulate a novel description of the
thermodynamics of solvable spin-$S$ $XXX$ models.
The suitable choices of auxiliary functions
yield a natural generalization of the strategy in \cite{KBP}.
The nonlinear integral equations close finitely,
which clearly differs from the string formulation\cite{Bab}
and is obviously efficient in numerics.
The resultant formulation has an interpretation in terms of
physical excitations, spinons and RSOS kinks.
This has been demonstrated by the calculation of the low-temperature
specific heat as well as the spinon character formula.
The latter, however, is derived under
several assumptions on the possible 
homotopy class of $L_{\cal C}$.
Certainly these constraints on winding numbers have 
"microscopic" origins in patterns of zeros of $y-$ (or $T$) functions.
This has been actually demonstrated for few cases:
the  superintegrable 3 state chiral Potts model \cite{KM} and 
the simplest case of
$sl_2$ RSOS models with open boundaries \cite{Melb}.
We hope to report on extensive numerical investigations on
zeros in the present context in the near future.

Finally we mention   spinon pictures 
 arisen from  different view points \cite{Schoutensp, BSlevel1, Hannover}.
where explicit "spinon" bases are constructed by vertex operators.
Thus the meaning of "spinon" is more 
transparent in comparison to the present approach.
Less obvious is their concrete relation to eigenstates of spin Hamiltonians.
The method also involves an uncontrolled approximation,
a truncation procedure,  
in evaluating  the partition function
as well as "one body" distribution functions.
This contrasts to the present approach which involves no approximation.
In a sense they are complementary, and their relations are to be explored.
%
%
\section*{Acknowledgments}
The author thanks A Kl{\"u}mper for discussions and 
critical reading of the manuscript.
He thanks A Fujii for discussions and helpful advice
on numerical calculations.
Thanks are also due to M Jimbo for explanation of his paper and
T Miwa for discussions and his interest in this paper.

\end{document}